\documentclass{amsart}
\usepackage{graphicx,tikz}
\usepackage[margin=1in]{geometry}
\usepackage[colorinlistoftodos]{todonotes}
\usepackage{color,nicefrac}
\newcommand{\D}{\mathcal{D}}

\usepackage[normalem]{ulem}

\begin{document}

\title[Partisan gerrymandering and the efficiency gap]{A formula goes to court:\\ Partisan gerrymandering and the efficiency gap}
\author{Mira Bernstein and Moon Duchin}
\thanks{Mira Bernstein
({\sf mira.bernstein@tufts.edu}) 
is a research assistant professor in Science, Technology, and Society
at Tufts University.  
Moon Duchin ({\sf moon.duchin@tufts.edu}) is an associate professor of 
Mathematics at Tufts University.}

\begin{abstract}
Recently, a proposal has been advanced to detect unconstitutional partisan gerrymandering
with a simple formula called the {\em efficiency gap}.  The efficiency gap is now working its way
towards a possible landmark case in the Supreme Court.  
This note explores some of its mathematical properties 
in light of the fact that it reduces to a straight proportional comparison of votes to seats.
Though we offer several critiques, we assess that 
$EG$ can still be a useful component of a courtroom analysis.  
But a famous formula can take on a life of its own and this one will need to be watched closely.
\end{abstract}
\maketitle

{\em Gerrymandering} is drawing political boundary lines with an ulterior motive.  This idea has 
great currency right now, with particular attention paid to manipulating shapes of U.S. 
congressional and legislative districts
in order to obtain a preferred outcome.
Gerrymandering comes in many flavors, including {\em racial gerrymandering}, where a minority 
group is subject to dilution of its voting strength; {\em partisan gerrymandering}, where 
one political party controls the process and tries to exaggerate its own political dominance; and {\em incumbent gerrymandering}, where
officials try to create safe seats for incumbents on both sides of the aisle.  
All kinds of gerrymandering use some of the same techniques, especially 
{\em packing}, where you seek to stuff your opponents into districts with very high percentages,
and {\em cracking}, where you disperse your opponents into several districts in numbers too small to 
predominate.  Both of these techniques generate wasted votes by your opponents and 
thus reduce your opponents' share of seats relative to their share of the votes cast.

In this note we will focus on partisan gerrymandering.
The Supreme Court has heard cases on partisan gerrymandering  three times,\footnote{{\em Davis v. Bandemer} (1986); {\em Vieth v. Jubelirer} (2004); {\em LULAC v. Perry} (2006)} and each time
the justices have balked.  They disagreed among themselves about whether partisan gerrymandering was within the Court's purview at all 
and about what standards might be used to detect it.  In the most recent case, the swing vote
belonged to Justice Anthony Kennedy, who wrote a lengthy opinion explaining why he was unsatisfied with the standards proposed by the plaintiffs to demonstrate that unconstitutional gerrymandering had occurred. Kennedy also gave some guidelines for what would be needed to satisfy him in the future. For the past 10 years, legal scholars and political scientists have pored over Kennedy's opinion, trying to design a standard that he might find convincing.

In 2015, one such team, Nicholas Stephanopoulos and Eric McGhee, advanced a new idea to quantify partisan gerrymandering, tailored to Kennedy's guidelines \cite{EG}. 
They propose a simple numerical score called the {\em efficiency gap} ($EG$) to
 detect and reject unfair congressional and legislative maps that are 
rigged to keep one party on top in a way that is unresponsive to voter preferences. $EG$ can be computed based on voting data from a single election: if the result exceeds a certain threshold, then the districting 
plan has been found to have discriminatory partisan effect.

Last November, for the first time in 30 years, a federal court  invalidated a legislative map as an unconstitutional partisan gerrymander \cite{Wisc}.\footnote{This happened only once before, in {\em Davis v. Bandemer} (1982), which was later overturned by the Supreme Court.}
A centerpiece in the district
court's ruling was the high efficiency gap in Wiconsin's 2012--2016 elections in favor of Republicans. 
Now the case is under appeal to the Supreme Court, and many observers are hoping that $EG$ will finally provide a legally manageable standard for 
detecting partisan gerrymanders in time for the 2020 census. If this happens, it will signal a seismic shift in American politics.

In this note, we engage in the following exercise:  first, we give a self-contained analysis for 
a mathematical audience, describing what it measures and what it does not.
Then, we examine how our findings relate to the original proposal, the press coverage, and the court
decisions to date.  We close by looking to the future career of this consequential formula.

\section{What is the efficiency gap?}
The U.S. electoral scene is dominated by the use of geographically-defined districts in which one
representative is chosen by a plurality vote. 
We will follow the $EG$ literature by simplifying to the case 
of only two political parties, $A$ and $B$. Let's say that our state has $S$ congressional or legislative seats, and denote the set of districts by
$\D=\{d_1,\dots,d_S\}$.  
In a particular election, write $\D=\D^A\sqcup \D^B$ where
$\D^P$ is the subset of districts won by party $P\in \{A,B\}$.
In what follows, if a value $X_i^P$ is defined with respect to party $P$ and district $d_i$, 
we will write $$X^P=\sum_{i=1}^S X_i^P \quad ; \qquad X_i=X_i^A+X_i^B \quad ; \qquad {\rm and}  \quad
X=X^A+X^B=\sum_{i=1}^S X_i.$$  

 Let $S_i^P$ be $1$ if party $P$ won in district $d_i$ and $0$ if not, so that $S^P=|\D^P|$ is the number of seats won statewide by $P$. Let $T_i^P$ be the number of votes cast in district $d_i$ for party $P$, so that 
$T_i$ is the  voting turnout in district $d_i$ and $T^A$ is the 
total number of statewide votes for $A$. Later we'll make the further simplifying assumption  that the districts have not just equal population but equal turnout: $T_i=T/S$.

With this notation, we can write $\tau=\frac{T^A-T^B}T$ and $\sigma=\frac{S^A-S^B}S$
for the statewide {\em vote lean} favoring $A$ and the statewide {\em seat lean} favoring $A$.
Most people's intuitions about fairness would incline towards a districting plan that 
approximates {\em  proportionality}, or $\sigma\approx\tau$.  Historically, the courts have sometimes
recognized
 rough proportionality as a virtue, but have explicitly rejected strict proportionality as a standard.\footnote{The 
 Supreme Court has a complicated attitude to proportionality.  
 See, for instance, {\em Gaffney v. Cummings} (1973), 
 {\em Davis v. Bandemer} (1986), and {\em Johnson v. De Grandy} (1994).  However, it was clearly stated in the {\em Bandemer}
 plurality opinion that  ``the mere lack of proportional representation will not be sufficient to prove unconstitutional discrimination." }

{\em Wasted votes}, in the $EG$ formulation, are any votes cast for the losing side 
or votes cast for the winning side in excess of  the 50\% needed to win.  
That is, the number of votes wasted by  $A$-voters in district $d_i$ is
$$
W_i^A \quad = \quad \begin{cases} T_i^A - \frac{T_i}2, & d_i\in \D^A, \\T_i^A, & d_i \in \D^B \end{cases} 
\qquad =\quad T_i^A - S_i^A\cdot\frac{T_i}2 
$$

Quick observation:  the total number of wasted votes in a district, 
$W_i=W_i^A+W_i^B$, is always half of the turnout $T_i$.  The question is how the wasted
votes are distributed.  If (nearly) all the wasted votes belong to the winning side, it's a packed district.  
If (nearly) all the wasted votes belong to the losing side, it's a competitive district.  And if there are several adjacent districts 
where most of the wasted votes are on the losing side, then it may be a cracked plan.

We can now define the {\em efficiency gap} associated with districting plan $\D$: 
$$
EG \;=\; \sum_{i=1}^S \frac{W_i^A-W_i^B}{T} \;=\; \frac{W^A-W^B}{T}.
$$
Thus $EG$ is a signed measure of how much 
more vote share is wasted by supporters of party $A$ than $B$.  
If $EG$ is large and positive, the districting plan is deemed unfair to $A$. On the other hand, 
$EG \approx 0$ indicates a fair plan, inasmuch as the two parties waste about an equal number of votes.
Stephanopoulos--McGhee write that this definition  ``captures, in a single tidy number, all of the packing and cracking decisions that go into a district plan."

For a toy example, consider the figure below, 
which shows three possible districting plans for the same distribution of voters. Each box represents a voter; $A$-voters are marked  $\star$ and $B$-voters are blank. Since $A$-voters make up 
half of the population, intuitively we expect a ``fair" plan to have $S^A = S^B$.  
Plans I and III 
both do this and it is easy to check that they have $EG=0$. In contrast, Plan II gives $A$ five of the six seats by packing some
 $B$-voters into one district and cracking the rest. This gerrymander is successfully detected: $EG = -1/3$, 
 unfairly favoring Party $A$.
 Note that there is a lot of packing and cracking in Plan III as well, 
 but this is not penalized by $EG$ because it happens symmetrically to voters of both parties. 

\vspace{.1in}

\begin{tikzpicture}[scale=.65]

\begin{scope}[xshift=-9cm]
\foreach \x in {0,...,6} {\draw [gray!85] (\x,0)--(\x,7);}
\foreach \y in {0,...,7} {\draw [gray!85] (0,\y)--(6,\y);}
\draw [rounded corners, line width=2.8, red] (0,7)--(4,7)--(4,6)--(3,6)--(3,5)--(0,5)--cycle;
\draw [rounded corners, line width=2.8, red, fill=red!20, fill opacity=.5] (4,7)--(6,7)--(6,4)--(4,4)--(4,5)--(3,5)--(3,6)--(4,6)--cycle;
\draw [rounded corners, line width=2.8, red] (0,5)--(4,5)--(4,4)--(3,4)--(3,3)--(0,3)--cycle;
\draw [rounded corners, line width=2.8, red] (3,4)--(6,4)--(6,2)--(2,2)--(2,3)--(3,3)--cycle;
\draw [rounded corners, line width=2.8, red, fill=red!20, fill opacity=.5] (0,3)--(2,3)--(2,2)--(3,2)--(3,1)--(2,1)--(2,0)--(0,0)--cycle;
\draw [rounded corners, line width=2.8, red, fill=red!20, fill opacity=.5] (3,2)--(6,2)--(6,0)--(2,0)--(2,1)--(3,1)--cycle;
\foreach \p/\q in {1/1,1/2,1/3,1/4,1/5,1/6,1/7,2/1,4/1,5/1,6/1,6/2, 6/3,6/5,6/6,6/7,5/7,4/7,4/5,4/4,5/3}  {\node at (\p-.5,\q-.5) {$\star$};}
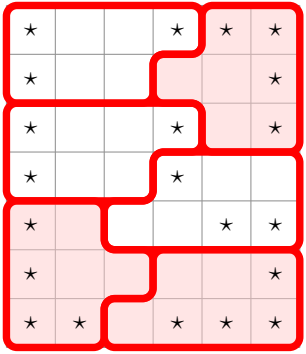
\node at (3,0) [below=.2cm] {Plan I: $EG=0$};

\end{scope}

\foreach \x in {0,...,6} {\draw [gray!85] (\x,0)--(\x,7);}
\foreach \y in {0,...,7} {\draw [gray!85] (0,\y)--(6,\y);}
\draw [rounded corners, line width=2.8, green!65!blue, fill=green!65!blue, fill opacity=.2] (1,7)--(6,7)--(6,5)--(5,5)--(5,6)--(2,6)--(2,5)--(1,5)--cycle;
\draw [rounded corners, line width=2.8, green!65!blue, fill=green!65!blue, fill opacity=.2] (0,7)--(1,7)--(1,5)--(2,5)--(2,2)--(1,2)--(1,3)--(0,3)--cycle;
\draw [rounded corners, line width=2.8, green!65!blue, fill=green!65!blue, fill opacity=.2] (3,5)--(4,5)--(4,4)--(6,4)--(6,2)--(3,2)--cycle;
\draw [rounded corners, line width=2.8, green!65!blue, fill=green!65!blue, fill opacity=.2] (0,3)--(1,3)--(1,2)--(2,2)--(2,3)--(3,3)--(3,1)--(2,1)--(2,0)--(0,0)--cycle;
\draw [rounded corners, line width=2.8, green!65!blue, fill=green!65!blue, fill opacity=.2] (3,2)--(6,2)--(6,0)--(2,0)--(2,1)--(3,1)--(3,2)--cycle;
\draw[rounded corners, line width=2.8, green!65!blue] (2,6)--(5,6)--(5,5)--(6,5)--(6,4)--(4,4)--(4,5)--(3,5)--(3,3)--(2,3)--cycle;
\foreach \p/\q in {1/1,1/2,1/3,1/4,1/5,1/6,1/7,2/1,4/1,5/1,6/1,6/2, 6/3,6/5,6/6,6/7,5/7,4/7,4/5,4/4,5/3}  {\node at (\p-.5,\q-.5) {$\star$};}
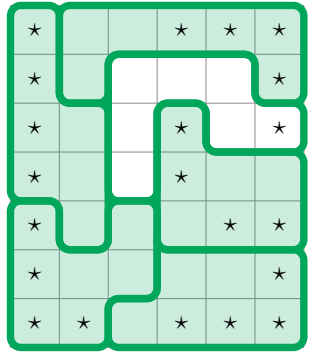
\node at (3,0) [below=.2cm] {Plan II: $EG=-1/3$};

\begin{scope}[xshift=9cm]
\foreach \x in {0,...,6} {\draw [gray!85] (\x,0)--(\x,7);}
\foreach \y in {0,...,7} {\draw [gray!85] (0,\y)--(6,\y);}
\draw [rounded corners, line width=2.8, blue, fill=blue!20, fill opacity=.5] (0,7)--(1,7)--(1,0)--(0,0)--cycle;
\draw [rounded corners, line width=2.8,  blue] (1,7)--(3,7)--(3,6)--(2,6)--(2,1)--(1,1)--cycle;
\draw [rounded corners, line width=2.8,  blue, fill=blue!20, fill opacity=.5] (3,7)--(6,7)--(6,4)--(3,4)--(3,5)--(5,5)--(5,6)--(3,6)--cycle;
\draw [rounded corners, line width=2.8,  blue] (2,6)--(5,6)--(5,5)--(3,5)--(3,4)--(4,4)--(4,2)--(3,2)--(3,3)--(2,3)--cycle;
\draw [rounded corners, line width=2.8,  blue] (2,3)--(3,3)--(3,2)--(4,2)--(4,4)--(6,4)--(6,3)--(5,3)--(5,1)--(2,1)--cycle;
\draw [rounded corners, line width=2.8,  blue, fill=blue!20, fill opacity=.5] (1,1)--(5,1)--(5,3)--(6,3)--(6,0)--(1,0)--cycle;
\foreach \p/\q in {1/1,1/2,1/3,1/4,1/5,1/6,1/7,2/1,4/1,5/1,6/1,6/2, 6/3,6/5,6/6,6/7,5/7,4/7,4/5,4/4,5/3}  {\node at (\p-.5,\q-.5) {$\star$};}
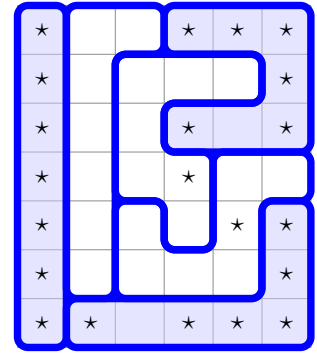
\node at (3,0) [below=.2cm] {Plan III: $EG=0$};
\end{scope}

\end{tikzpicture}

\vspace{.1in}

Of course, in real life, no districting plan will have an efficiency gap of exactly 0. How high is too high? In \cite{EG}, the authors argue that  $EG=.08$ corresponds to a historically 
robust threshold for unacceptable partisan gerrymandering.\footnote{They propose a somewhat
different standard of {\em two excess seats} for congressional districting plans.\cite{EG}} In the  Wisconsin case, for example, the plaintiffs demonstrated that the last three elections for the state assembly had efficiency gaps between $0.1$ and $0.13$ in favor of Republicans.  Since it was a Republican legislature that had drawn the map, and there was plentiful evidence that they had intentionally done so to disfavor Democrats, the court ruled that the plan was an unconstitutional partisan gerrymander. QED.

\section{What does the efficiency gap actually measure?}

But wait!  
The $EG$ formula turns out to simplify quite a bit, in a way that has bearing on our understanding of 
what it measures.
Let's proceed with the assumption of
 equal voter turnout: $T_i = T/S$.\footnote{Dropping the equal-turnout assumption only makes matters
worse in the critiques of $EG$ that follow, because this corresponds to a weighting of terms that is harder to interpret and defend.  Without the equal-turnout simplification, $EG$ is affected as follows:
if there is lower average turnout in the districts $\mathcal D^A$, then maintaining a low $EG$ requires
$A$ to get more seats than party $B$ would have with the same vote share.}
In this case, we get
$$
W^A \;=\; \sum W_i^A \;=\; T^A-S^A\cdot\frac{T}{2S},
$$
and thus
$$
EG \;\;=\;\;  \frac{T^A-T^B}{T} \;-\; \frac 12 \frac{S^A-S^B}{S} \;\;=\;\; \tau-\frac 12 \sigma.
$$
\smallskip

That is:  the efficiency gap is just the statewide vote lean favoring $A$ minus half of the 
statewide seat lean favoring $A$.%
\footnote{This can also be written in terms of 
{\em seat share} and {\em vote share} rather than seat lean and vote lean:\\
$EG=2(T^A/T)-(S^A/S)-1/2.$}
It has nothing at all to do with how the voters are distributed among districts, per se.  As long as the seat total
comes out a certain way, as in Plan III shown above, $EG$
 does not penalize packing or cracking, or for that matter bizarrely-shaped districts---and indeed, it sometimes 
incentivizes them, as we will see below.

\vspace{.1in}

In its simplified form $(\tau-\frac 12 \sigma)$, we can see that $EG$ has numerous potentially undesirable 
properties.

\smallskip

\begin{itemize}
\item {\bf Penalizes proportionality.}
If Party $A$ has 60\% of the statewide vote and 60\% of the seats, 
$EG$ rates this as an unacceptable gerrymander in favor of Party $B$!  ($\tau - \frac{\sigma}2=0.2-0.1>0.08.$)
This is because $EG=0 \iff \sigma=2\tau$.  That is, the intuitive idea that 
representation should be proportional to vote share is replaced by the conflicting 
principle that the seat lean should be {\em twice} the vote lean.

\smallskip

\item {\bf Volatile in competitive races.}
$EG$ behaves very erratically if there are districts with competitive races, because a genuinely
close outcome will produce  lopsided vote wastage, but it is unpredictable which side this falls 
on.\footnote{In a current preprint, Cover argues persuasively that this provides a stark disincentive
for even an honest mapmaker to draw a competitive plan \cite[p34]{Cover}.}
If, for instance, all districts are competitive but a last-minute trend pushes voters to one side
systematically, then the {\em plan itself} will be rated as a gerrymander.

\smallskip

\item {\bf Fetishizes three-to-one landslide districts.}   We've seen that $EG$ is not sensitive to 
any changes in packing and cracking that preserve $\sigma$ (i.e., that preserve the overall seat outcome).
But if anything, $EG$ rewards a certain level of district-by-district packing.
Recall that every district has a total of 50\% vote wastage.  It immediately
follows that the only way to share that fairly in a single district is to have 25\% on each side, which is 
a 75-25 vote split.  So the only districts viewed by $EG$ to be perfectly neutral 
are highly non-competitive districts, and any plan made up entirely of these landslide
districts will be judged perfectly fair.

\smallskip

\item {\bf Breaks down in edge cases.}
Despite the fact that $75\%$ is an artificial sweet spot, $80\%$ statewide vote share breaks $EG$ completely.
If $A$ controls more than 79\% of the vote in a state,  then $\tau > .79 - .21 = .58$. In order to get $\tau-\sigma/2<.08$, we must have $\sigma>1$, which is impossible.
Thus, in this circumstance 
$EG$ will identify absolutely {\em any} districting plan as a partisan gerrymander in favor of $B$. 

\smallskip

\item {\bf Nongranular.}
We have seen that $EG$ does no more and no less than compare seats to votes
by a double-proportionality standard.   
This has the added consequence that in states with a small number of districts, the way that $EG$
depends on a districting plan is extremely nongranular:  for a given vote split, there are only
$S$ possible values of the the efficiency gap, as the majority party's seat total ranges 
from 1 to $S$.  For a particular voting split, a small state may have no outcome at all with a 
permissibly small $EG$. 
This makes the score far too coarse for use in many states; 21 states currently
have four or fewer Congressional districts.%
\footnote{Cho discusses nongranularity at length in  \cite{Cho}.}

\end{itemize}

\section{How is this playing out in court?}

So far we have  highlighted, with the help of  simple algebra, several grave limitations 
of $EG$ as a stand-alone metric.
If the Supreme Court were to enshrine the
efficiency gap as a dispositive indicator of partisan gerrymandering, it would
be sure to produce  false positives as well as false negatives with respect to any common-sense
understanding of political unfairness.
But has anyone actually proposed that it be used that way?  
The popular press has touted $EG$ as ``a formula to end gerrymandering"; 
a simple way to ``end partisan gerrymandering once and for all"; and trumpeted that 
 ``this is how to get rid of gerrymandered districts."  
But a brief review of the political science literature and court documents shows 
circumspect writing and multifaceted analysis at all stages:
the role of  $EG$ in the legal landscape is much more complex.

\subsection*{Original framing}

First and importantly, 
Stephanopoulos and McGhee propose a doctrinal test
for partisan gerrymandering in which $EG$ can  be considered only if several other conditions are 
met.
They address concerns about volatility 
by requiring evidence of stability:  plaintiffs must perform 
a ``sensitivity analysis" to show that their disadvantage would persist under a modest statewide swing to one party or the other and must demonstrate the gap's likely durability over time.
Nongranularity is indirectly acknowledged:  in their historical analyses, the authors
only considered elections with eight or more seats. 
And edge cases, in the form of extremely lopsided elections, are simply dismissed as 
never occurring at the statewide level.

Defending the idealization of double-proportionality is a heavier lift.
The fact that using $EG$ commits you to condemning proportional outcomes 
is actually observed in a footnote \cite[p18 note 107]{EG}, but the authors defend it as 
reasonable on the grounds of quantifying an appropriate ``seat bonus" for the majority party. On this view, 
the efficiency gap quantifies the level of bonus that {\em should} be enjoyed by the winner:
a party with $60\%$ of the vote should have 70\% of the seats, a party with $70\%$ of the vote should have $90\%$ of the seats, etc. We will return to this point below.


\subsection*{The district court}

Many objections to the usefulness of $EG$ were
introduced in the Wisconsin case in the form of expert reports and testimony for the defense.  
The court decision 
endorses a three-pronged test\footnote{First, plaintiffs must establish the {\em intent} to 
discriminate on the basis of political affiliation.
Then, discriminatory {\em effect} must be established, and $EG$ can be used to that end.  And finally, defendants must fail
to provide {\em justification} of the necessity of the plan on other legitimate legislative grounds.}  
expanding on  \cite{EG} by requiring proof of discriminatory intent,
 and the majority opinion argues that this
protects against false positives:
``If a nonpartisan or bipartisan plan displays a high $EG$, the remaining components of the analysis will prevent a finding of a constitutional violation."  Thus, for instance, a proportional plan would not be thrown 
out on the grounds of high $EG$, because the plaintiffs would not be able to demonstrate improper
partisan intent.
The court affirms $EG$ not as a conclusive indicator, but only as persuasive ``corroborative evidence of an aggressive partisan gerrymander."

\subsection*{The Supreme Court}

The strategy that the Wisconsin plaintiffs will use at the next level is outlined in their motion to appeal.  
While praising the usefulness and diminishing the critiques of $EG$, this document does emphasize its limited role:
``To be clear, Appellees do not ask the Court to endorse any particular measure of partisan asymmetry or any particular technique for demonstrating durability. The [district court] did not do so, nor need the [Supreme Court] in order to affirm. Rather, Appellees advocate the same course of action the Court has followed in other redistricting contexts involving discriminatory effects: namely, the articulation of a standard whose precise contours are filled in through subsequent litigation."  

\subsection*{And beyond}

There is a philosophical issue at the heart of this analysis.  
Can a formula be said to ``measure" quantities that are used to compute it, or only those 
to which it is numerically sensitive?  Wasted votes are apparent inputs into $EG$, but their local 
contributions come out in the wash, 
and only {\em deviation from double-proportionality of seats to votes} remains. 
Setting double-proportionality as a target contravenes the common-sense preference for proportionality,
an ideal that even the Supreme Court has recognized despite its insistence that it is not constitutionally
required.
While political scientists have found evidence of a  hyper-proportional seat bonus 
in real elections, we have seen no persuasive case for the slippage from an empirical description to
a normative standard (that is, from what {\em is} the case to what {\em should be} the case).  
We assess this as the most serious indictment of the $EG$ formula.  

Our evaluation suggests a suite of particular circumstances in which 
$EG$ could be a useful component of a broader analysis\footnote{To use $EG$, we'd want
a state with enough 
congressional districts for $EG$ to be sufficiently granular, 
a close enough overall partisan preference split that double-proportionality
does not predict outlandish seat margins, few enough uncontested 
districts to require a minimum of counterfactual speculation in incorporating them in to the numbers,
an appropriate balance of competitive districts to pass sensitivity analysis, and
 an egregious enough partisan power grab to still show up.}---and it seems that 
Wisconsin is currently such a case.  
However, a mathematician could well argue that a formula which is only used to lend numerical corroboration in special and extreme cases is a formula of limited usefulness indeed.  
This caution has to be weighed against the plain fact that courts have found no manageable standard
to date for even the most extreme partisan abuses in redistricting.  

To a great extent, 
 critiques of $EG$ are mitigated by the limited circumstances in which courts will apply 
it and the fact that it may well 
be endorsed only as a first draft of a general standard that will be refined over time.
But major concerns remain.
Courts will have safeguards in place, but $EG$ is already in play outside of courts,
and there is a real risk that it may come to stand in as an operationalized {\em definition} of partisan gerrymandering.
We are  seeing hints of this effect not only in the overheated popular press coverage 
but in more scholarly work as well.\footnote{One such example is {\em Extreme Maps}, a research report put out by the Brennan Center at NYU, which uses $EG$ without any caveats 
to assess the current state of partisan gerrymandering in the U.S.}

Legal scholars believe that $EG$ will appeal to the courts because of its simple, one-shot construction with 
no technical machinery. As we have seen, the simplicity is actually illusory: a lot of care, including further statistical testing and modeling, is required to use $EG$ responsibly. 
Moreover, $EG$ comes on the scene at a time when having a single formula is becoming less important. 
One of the most promising directions in the detection of gerrymandering is the use of 
 supercomputing algorithms that can take multiple  indicators into account simultaneously.
For instance, various teams of researchers have developed simulation tools for comparing a proposed plan 
against an algorithmically generated batch of alternate plans.\footnote{See for instance multiple papers 
with various co-authors by Jowei Chen and by Jonathan Mattingly, whose teams use 
 Markov chain Monte Carlo, or by Wendy Cho, whose team uses genetic and evolutionary algorithms.}  
Typically these simulations can incorporate both equality constraints and inequality constraints, encoding
both legal requirements and preferences.  
The core idea of $EG$ can be coded into a simulation analysis by, 
for instance, replacing the use of $\tau-\frac 12 \sigma$ 
as a score  with bounds that 
constrain the deviation from proportionality rather than prescribing it.  
Even better, the seats-to-votes proportionality could be used as an evaluation axis.  A plan like the Wisconsin 
legislative map could be taken as input, and hundreds of thousands of alternate maps would 
be randomly generated that meet the relevant legal requirements and are scored at least as well by legitimate
districting criteria.  If a districting plan  had a seat share to vote share ratio
of $\nicefrac{({S^A}/{S})}{({T^A}/{T})}=1.25$ in a particular election and 
$95\%$ of computer-generated alternatives had ratios of $1.05$--$1.15$ with the same data, 
then we'd have excellent
evidence of excessive partisan skew.  
Use of simulation comparisons is improving quickly as increased computing power lets the algorithms  
visit more of the space of possible plans.   As the generation of random alternative maps rapidly becomes
both sophisticated and practicable,  the hoped-for breakthrough in 
adjudicating gerrymanders, partisan and otherwise, may be coming within close reach after all.

The Wisconsin plaintiffs are not asking the court to enshrine $EG$ as the one true measure of
partisan gerrymandering, but only to accept it as a starting point in building a test to show when 
entrenched partisan advantage has risen to the level of vote dilution of political opponents.
We hope that the Supreme
Court agrees with them in a decision that leaves room for $EG$ to pave the way for refined metrics
and methods in the years to come.


\begin{thebibliography}{99}
\bibitem{Cho} Wendy K. Tam Cho,  
{\em Measuring Partisan Fairness: Guarding Against Sophisticated as well as Simple-Minded Modes of Partisan Discrimination},  
University of Pennsylvania Law Review Online, {\bf 166} (July 2017). 
\bibitem{Cover} Benjamin Plener Cover, {\em Quantifying Political Gerrymandering: An Evaluation of the Efficiency Gap Proposal}, 
70 Stanford Law Review (forthcoming 2018).
\bibitem{EG}
Stephanopoulos, N. O., \& McGhee, E. M. (2015). {\em Partisan gerrymandering and the efficiency gap.}  
The University of Chicago Law Review, 831--900.
\bibitem{Wisc} Whitford v. Gill, F.Supp. 3d, 2016 WL 6837229, 15-cv-421-bbc, W.D.Wisc. (Nov. 21, 2016).
\end{thebibliography}
\end{document}